\documentclass[twocolumn]{aastex631}
\usepackage[T1]{fontenc}
\usepackage{savesym}
\savesymbol{tablenum}
\usepackage[section]{placeins}
\usepackage{amsmath} 
\usepackage{siunitx} 
\restoresymbol{SIX}{tablenum}
\usepackage{amssymb}
\usepackage{natbib}
\usepackage{changepage}
\usepackage{graphicx}
\usepackage{ragged2e}
\usepackage{threeparttable}
\justifying
\usepackage{url}
\usepackage{xcolor}

\shorttitle{Ejective feedback in a high-z quiescent galaxy}
\shortauthors{Wu}

\graphicspath{{./}{figures/}}

\begin{document}

\title{Ejective feedback as a quenching mechanism in the 1.5 billion years of the universe: detection of neutral gas outflow in a $z=4$ recently quenched galaxy}

\author{Po-Feng Wu}
\affiliation{Institute of Astrophysics, National Taiwan University, Taipei 10617, Taiwan}
\affiliation{Department of Physics and Center for Theoretical Physics, National Taiwan University, Taipei 10617, Taiwan}
\affiliation{Physics Division, National Center for Theoretical Sciences, Taipei 10617, Taiwan}
    
\begin{abstract}
The confirmation of massive quiescent galaxies emerging within the first billion years of the universe poses intriguing questions about the mechanisms of galaxy formation. There must be highly efficient processes at work to shut down star formation in galaxies at cosmic dawn. I present the detection of neutral outflowing gas in a massive recently quenched galaxy at $z=4$, showing ejective back as a quenching mechanism. Based on JWST spectrum, the star formation rate of this has been declining with a rapid e-folding timescale of $\sim50$~Myrs. The current specific star formation rate is $5\times10^{-11}$~yr$^{-1}$, roughly 40 times lower than that of the star-forming main sequence at comparable redshifts. Emission line ratios of [\ion{Ne}{3}]/[\ion{O}{2}] and [\ion{O}{3}]/H$\beta$ are similar to AGN at comparable redshifts. A series of \ion{Fe}{2} and \ion{Mg}{2} absorption lines appear blueshifted by $\sim250$~km~s$^{-1}$ relative to the stellar continuum, suggesting an outflow of neutral gas. The estimated mass outflow rate is approximately 7 times greater than the star formation rate derived from the stellar continuum, implying that the suppression of star formation is likely due to gas being depleted by the outflow. If the emission lines are AGN-driven, the AGN can also provide sufficient energy to launch the outflow observed. This galaxy represents the most distant example of its kind known to date. This study offers a compelling explanation for the existence of massive quiescent galaxies in the first billion years of the universe. 

\end{abstract}

\section{Introduction} \label{sec:intro}
Large extra-galactic imaging surveys have revealed that the rest-frame optical colors of galaxies exhibit a bimodal distribution \citep{str01,bal04,bel04,wil06,fra07}. Galaxies with blue colors are typically those with active in-situ star formation, whereas red galaxies show minimal star formation activities. Based on redshifts and restframe colors derived from broadband and medium band spectral energy distribution (SED) fitting, a population of massive (several $10^{10} M_\odot$) red galaxies has been identified in the first two billion years of universe \citep{ilb13,muz13b,str14}. In addition to their rapid formation, what makes these galaxies even more intriguing is their early quiescence. Despite the overall increase in cosmic star formation rate density during this period \citep{mad14}, there must have been highly efficient mechanisms in the early universe that truncated star formation activities in these galaxies.

However, photometric SED fitting can introduce significant uncertainties in redshifts and restframe colors. Spectroscopic follow-up observations have demonstrated that many galaxies initially identified as massive with $z_{phot}>3$ are in fact lower-redshift dusty star-forming galaxies \citep{for24}. Nevertheless, deep near-IR spectroscopic observations have confirmed the existence of dozens of massive quiescent galaxies at $z>3$ \citep{mar15,sch18,for20a,deu21,for22,mcc22,tan24,set24}, highlighting a population that current numerical simulations struggle to reproduce, especially the most massive ones \citep{for20b}. 

Furthermore, several massive quiescent galaxies at $z>4$ have also been spectroscopically confirmed, both through their faint emission lines and stellar populations, as constrained by stellar continuum absorption \citep{tan19,car23,kak24,deg24,fry24,car24}. Full-spectral fitting of the stellar continua indicates that most of these galaxies build up the bulk of their stellar masses during a brief but intense period of star formation. During this phase, the star-formation rate (SFR) peaks at least a few hundred $M_\odot$~yr$^{-1}$, followed by a rapid truncation with an exponential decaying timescale on the order of $\sim100$~Myr. This violent process is not a typical mode of star formation at $z\sim4$, where the gas depletion time through star formation, $M_{gas}/SFR$, is generally several hundred Myrs for most star-forming galaxies \citep{des20}. Therefore, the abrupt decline in SFR in these quiescent galaxies must have been triggered by additional processes. 

Ejective feedback, driven by either active galactic nuclei (AGNs) or intense starbursts, is widely considered a key mechanism for the abrupt truncation of star formation in galaxies. The immense energy and momentum released in these events can propel powerful outflows, expelling gas from the galaxy and depleting the gas reservoir far more rapidly, effectively halting star formation on much shorter timescales  \citep{dim05}. 

High-velocity gas outflows have been detected in quiescent galaxies that recently experienced a sharp decline in star formation rates (SFRs). These recently quenched galaxies, often referred to as post-starburst galaxies, offer valuable insights into quenching mechanisms. Notably, observations of blueshifted \ion{Mg}{2} absorption with velocities exceeding 1,000 km s$^{-1}$ have been reported in several recently quenched galaxies at $z < 1.5$ \citep{tre07,coi11,mal19,tay24}, indicating outflowing neutral gas. Some but not all of these galaxies host AGN. Moreover, \citet{wu23} found spatially extended blueshifted, AGN-driven [\ion{O}{3}] gas in recently quenched galaxies at $z\sim0.7$, illustrating the multi-phase nature of gas outflows in these systems. In contrast, most recently quenched galaxies show no evident signs of outflow in their molecular gas phase \citep{wu23,zan23}. 

However, these studies have not reported the mass outflowing rates. \ion{Mg}{2} absorption is typically optically thick, offering only a lower limit on the gas column density. Meanwhile, [\ion{O}{3}] emission traces the ionized gas, which constitutes a minor faction of the overall mass budget \citep{fio17,ave22}. Thus, while these outflows are observed, it remains unclear whether they significantly impact star formation. 

James Webb Space Telescope (JWST) has recently enabled the measurement of interstellar medium (ISM) absorption in massive quiescent galaxies at higher redshifts, thanks to its extended wavelength coverage and enhanced sensitivities. \citet{deu23} and \citet{bel24} reported the detection of blueshifted \ion{Na}{1} absorption of two young quiescent AGN-host galaxies at $z\simeq2.5$ and $z\simeq3$. Based on the \ion{Na}{1} absorption strengths, their outflow rates of neutral gas surpass their SFRs by an order of magnitude. These outflows are depleting gas reservoirs more rapidly than star formation. These findings are claimed to be a smoking gun for ejective feedback, especially from supermassive black holes (SMBHs), potentially explaining the presence of massive quiescent galaxies at higher redshifts.

In this paper, I push the redshift frontier by presenting new evidence of ejective feedback as a quenching mechanism beyond $z>4$. I discuss the detection of neutral gas outflow and the associated outflow rate, traced by \ion{Mg}{2} and \ion{Fe}{2} absorption lines, in a massive, recently quenched galaxy at $z=4.1$. This galaxy currently stands as the highest-redshift example of its kind, offering crucial insight into the existence of massive quiescent galaxies at $z>4$.

In Section~\ref{sec:target}, I introduce the target galaxy and observations conducted. Section~\ref{sec:opt} describes the modeling of restframe optical spectra for the star-formation history (SFH) and ionized line emissions. In Section~\ref{sec:outflow}, I calculate the physical properties of the neutral gas outflow. The implications of these findings are discussed in Section~\ref{sec:dis}, and the key takeaways are summarized in Section~\ref{sec:sum}.

\section{The target}
\label{sec:target}

\begin{figure*}
    \centering 
    \includegraphics[width=0.95\textwidth]{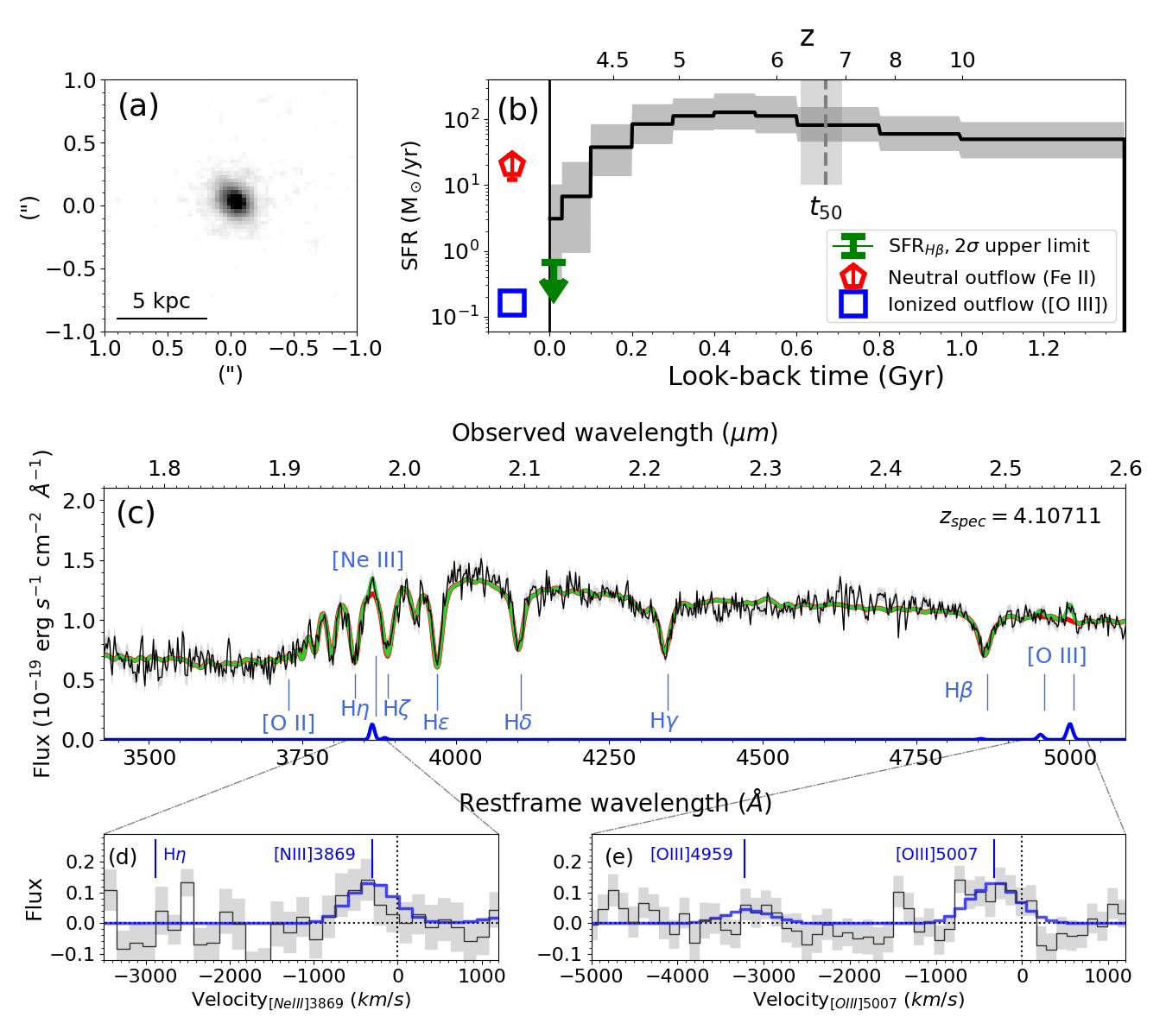}
    \caption{(a) The $2"\times2"$ F200W cutout image of the galaxy.  (b) The SFH of the galaxy constrained by the stellar continuum. The black line and the shaded region represent the median and the 16th and 84th percentiles of the posterior distributions, respectively. The galaxy formed half of its stellar mass at $z\sim6.7$ (vertical dashed line). The SFR peaked at a lookback time of $\sim400$~Myr and declined from $\sim130\ M_\odot \mbox{yr}^{-1}$ to $\sim3\ M_\odot \mbox{yr}^{-1}$. The 2-$sigma$ upper limit of the SFR derived from the H$\beta$ emission line flux (green arrow) is also well below the star-formation main sequence (Section~\ref{sec:opt}). The red pentagon and blue square represent the outflow rates of neutral and ionized gas, respectively (Section~\ref{sec:outflow}). The error bar for the neutral outflow reflects uncertainty from the column density, while the error for the ionized outflow, constrained by flux measurements, is smaller than the label. Systematic uncertainties are not included in the error bar (see Section~\ref{sec:outflow}). The quantities are all measured from flux in the slit and not corrected for the 50\% slit loss. (c) The restframe optical spectrum of the galaxy (black) and the best-fit models. The blue line represents the emission line component derived from the median of 500 pPXF fits (see text), while the red line shows the stellar continuum model from BAGPIPES. The green line denotes the sum of the red and the blue components. Emission lines included in the pPXF are labeled. The spectrum indicates a low sSFR as no prominent star formation indicators, such as [\ion{O}{2}] and H$\beta$, are detected with high significance. (d,e) Zoom in to the detected emission lines. [\ion{Ne}{3}]$\lambda$3869 and [\ion{O}{3}]$\lambda$5007 are the only 2 lines with $S/N>2$. \label{fig:gal}}
\end{figure*}

The galaxy discussed in this paper was first reported by \citet{fry24} as a quiescent galaxy with spectroscopic redshift $z = 4.1076\pm0.0023$ (NS\_274 in Table~1 of \citet{fry24}, hereafter, SNH0pe-NS274). This observation was conducted as part of the JWST Director's Discretionary Time (DDT) program (PID: 4446, PI: B. Frye). The spectra were obtained using the NIRSpec Micro-Shutter Array (MSA) with the grating/filter combinations G140M/F100LP and G235M/F170LP, providing a spectral resolution $R\sim1000$. The on-source exposure times are 4420 seconds for G140M/F100LP and 6996 seconds for G235M/F170LP. For further details on the observation setting, readers are referred to \citet{fry24}.

I use the 1D spectra from the DAWN JWST Archive (DJA)\footnote{\url{https://dawn-cph.github.io/dja/index.html}}, which is reduced with \texttt{msaexp} v.0.6.11 \citep{bra23}. \texttt{msaexp} uses the Stage 2 output from the MAST JWST archive. The wavelength, flat-field, and photometric calibrations of individual exposure files were done with relevant reference files associated with the Calibration Reference Data System (CRDS) context \texttt{jwst\_1084.pmap}. The bias levels and the $1/f$ noise are corrected for.

Figure~\ref{fig:gal}a shows the image in NIRCam/F200W from the same JWST program. The image is Stage 3 product from MAST JWST archive.\footnote{The data is available at MAST: \dataset[10.17909/s0jn-ys04]{\doi{10.17909/s0jn-ys04}}. } I use \texttt{galfit} \citep{pen10} to fit a single S\'{e}rsic profile to the 2D light distribution. The point-spread function (PSF) is determined by stacking images of bright stars. According to the best-fit model, the galaxy has an AB magnitude of $m_{AB,F200W} = 23.1$, an effective radius $R_e = 390$~pc, a S\'{e}rsic index $n=5.3$, and an axis ratio of 0.6. This analysis reveals that the galaxy is a compact, flattened spheroid. 

Convolving the spectrum (Figure~\ref{fig:gal}b) with the transmission curve of the F200W filter, I measure the flux through the slit is 50\% of the flux measured from the F200W image, indicating a 50\% slit loss. From the slit spectrum, I derive a stellar mass of $M_\ast = 6 \times 10^{10} M_\odot$ (see Section~\ref{sec:pipes} and Table~\ref{tab:pipes_param}). Assuming the mass-to-light ratios are the same inside and outside the slit, the total mass of this galaxy is $M_\ast = 1.2 \times 10^{11} M_\odot$. The effective radius of this galaxy is broadly consistent with other quiescent galaxies \citep[$< 1$~kpc, ][]{tan19,car23,deg24} and is several times smaller than that of star-forming galaxies with similar stellar masses and redshifts \citep{war24}.

\section{Stellar population and ionized gas}
\label{sec:opt}

The G235M grism spectrum covers the restframe optical wavelengths, encompassing age-sensitive stellar absorption lines and prominent ISM emission lines (Figure~\ref{fig:gal}c). I use the spectrum between 3480 \AA\ and 5200 \AA\ in the restframe to constrain the stellar population and the star formation history (SFH) of the galaxy. Unless stated otherwise, the quantities reported are based on the fluxes measured through the slit, with no correction applied for the 50\% slit loss.

\subsection{Modeling the emission lines}
\label{sec:ppxf}

\begin{table}
\centering
\hspace*{-1.5cm}
\begin{threeparttable}
	\caption{Emission line properties}
\begin{tabular}{lc}
	\hline
	\hline
	Line & flux  \\
             & ($10^{-19}$ ergs cm$^{-2}$ s$^{-1}$ ) \\
	\hline  
	  [\ion{O}{2}]$\lambda 3727,3729$ & 
 $0.02^{+0.22}_{-0.02}$ \\
 
        [\ion{Ne}{3}]$\lambda 3869$ & $1.16^{+0.28}_{-0.50}$ \\ 
        H$\beta$                  & $0.13^{+0.19}_{-0.13}$ \\
        
        [\ion{O}{3}]$\lambda 5007$ & $1.98^{+0.24}_{-0.65}$ \\ 
	\hline
        \hline
        Kinematics    & \\
        \hline
        $v$\tablenotemark{a} & $-307^{+45}_{-36}$~km~s$^{-1}$ \\
        $\sigma$ & $\quad 260^{+26}_{-24}$~km~s$^{-1}$ \\
        %v_{forbidden}$ & $-307^{+45}_{-36}$~km~s$^{-1}$ \\
        %$\sigma_{forbidden}$ & $\quad 260^{+26}_{-24}$~km~s$^{-1}$ \\
        %$v_{Balmer}$ & $-26^{+45}_{-56}$~km~s$^{-1}$ \\
        %$\sigma_{Balmer}$ & $\quad 62^{+46}_{-60}$~km~s$^{-1}$ \\
        \hline
\end{tabular}
\tablenotetext{a}{The velocity relative the the stars.}
\label{tab:em}
\end{threeparttable}
\end{table}

To determine the fluxes and velocities of these emission lines, I use the Penalized Pixel-Fitting method \citep[pPXF,][]{cap04,cap17,cap23}, which fits a model comprising both the stellar continuum and gas emission. The stellar template is based on the UV-extended E-MILES stellar population synthesis models \citep{vaz16}. All emission lines are modeled as a single kinematic component. Each line is described as a Gaussian and all lines have the same velocity and velocity dispersion. The strength of each line is treated as a free parameter, except [\ion{O}{3}], where the flux ratio of [\ion{O}{3}]$\lambda$5007/[\ion{O}{3}]$\lambda$4959 is fixed at 3. 

For the pPXF analysis, the input spectrum is de-redshifted using the redshift determined from the stellar continuum fit with BAGPIPES (Section~\ref{sec:pipes}). The resulting stellar velocity offset is within a few tens of km~s$^{-1}$, a sub-pixel difference that aligns well with the BAGPIPES redshift uncertainty.

To estimate the uncertainties, I perturb the spectrum by adding a random number drawn from a Gaussian distribution to each pixel, where the width and the strength of the Gaussian are determined by the error spectrum and then fit the spectrum. The operation is repeated 500 times. Table~\ref{tab:em} presents the 16th, 50th, and 84th percentiles of relevant line fluxes, line velocity dispersion, and line velocity relative to the stars from these 500 iterations. Only two emission lines have signal-to-noise ratios above 2: [\ion{Ne}{3}]$\lambda$3869 with $S/N = 2.2$ and [\ion{O}{3}]$\lambda$5007 with $S/N = 3.0$, while the fluxes for all other lines are consistent with 0.
The emission lines are blueshifted relative to the stellar continuum, with a velocity of $\sim$300~km~s$^{-1}$ (Figure~\ref{fig:gal}d,e). The [\ion{O}{2}]$\lambda$3727,3729 doublet, which is typically prominent and often observed in quiescent galaxies \citep{mas21}, is not detected. H$\beta$ is also not detected.

\begin{figure}
    \centering 
    \includegraphics[width=0.95\columnwidth]{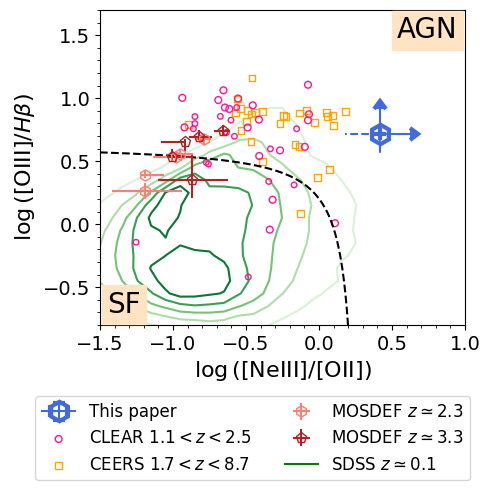}
    \caption{The [\ion{Ne}{3}]/[\ion{O}{2}] v.s. [\ion{O}{3}]/H$\beta$ line ratio of SNH0pe-NS274 (blue hexagon). I take the 2-$\sigma$ upper limit of [\ion{O}{2}] H$\beta$ to calculate the lower bound of [\ion{Ne}{3}]/[\ion{O}{2}] and \ion{O}{3}/H$\beta$. The lengths of the dashed lines are calculated using 1-$sigma$ lower bound of [\ion{Ne}{3}] and [\ion{O}{3}]. For comparison, several samples of emission-line galaxies from various projects are plotted: SDSS at $z\simeq0.1$ \citep[green contours,][]{aba09}, CLEAR at $1.1<z<2.5$ \citep[magenta circles,][]{sim23}, CEERS at $z>2$ \citep[yellow squares,][]{bac24}, and stacked spectra of MOSDEF at $z\sim2.3$ and $z\simeq3.3$ \citep[round and orange pentagons,][]{san21}. The black dashed line represents the boundary between star formation and AGN ionization \citep{bac22}. SNH0pe-NS274 exhibits among the highest [\ion{Ne}{3}]/[\ion{O}{2}] and [\ion{O}{3}]/H$\beta$ line ratios. \label{fig:ohno}}
\end{figure}

Figure~\ref{fig:ohno} presents a comparison of the line ratios of the target quiescent galaxy with those of emission-line galaxies from various studies across a broad redshift range: SDSS DR7 at $0.08 < z < 0.12$ \citep{aba09}, the CLEAR survey at $1.1 < z < 2.5$ \citep{sim23}, the CEERS survey at $z>1.7$ \citep{bac24}, and the stack spectra from galaxies in different stellar mass bins from the MOSDEF survey at $z\sim2.3$ and $z\sim3.3$ \citep{san21}. Only galaxies with $S/N > 2$ for all 4 lines are plotted. The dashed line in Figure~\ref{fig:ohno} delineates the boundary between AGN and \ion{H}{2} regions, as proposed by \citet{bac22}. 

Using a $2\sigma$ upper limit for [\ion{O}{2}] and H$\beta$, the lower of limit of [\ion{Ne}{3}]/[\ion{O}{2}] for SNH0pe-NS274 exceeds all other galaxies in the comparison. Along with its [\ion{O}{3}]/H$\beta$, these line ratios are similar to AGN at $z>5$ and suggest a high ionization parameter, with $\log[q/(cm\ s^{-1})] > 8 $  \citep[see discussion in][]{koc23,bac24}.

\subsection{Modeling the star formation history}
\label{sec:pipes}

I use the code BAGPIPES \citep{car18} to fit the spectrum and constrain the SFH, the redshift, and the stellar mass. The stellar continuum templates are generated with the 2016 revised version of the \citet{bc03} population synthesis model \citep{che16} using the E-MILES stellar spectral library \citep{fal11,vaz16}. The models are convolved with a Gaussian kernel in velocity space to account for the stellar velocity dispersion of the galaxy. 

The dust attenuation law follows the model of \citet{sal18}, which is parameterized as a power-law deviation from the Calzetti dust law \citep{cal20}. The \citet{sal18} dust law also incorporates a Drude profile to model the 2175\AA\ bump, though this feature is irrelevant here. For stars younger than 10~Myrs, a different attenuation is permitted and is parameterized as $\eta$, the ratio of attenuation in magnitudes between the young and the old populations. 

Based on the redshift reported by \citet{fry24}, I allow the redshift to vary within a narrow range $4.09 < z < 4.12$. I adopt a non-parametric SFH with 11 age bins, defined by bin edges at 0, 10, 30, 100, 200, 300, 400, 500, 600, 800, 1000~Myrs, and the age of the universe. In this model, the SFR within each age bin is assumed to be constant. The model fits for $\Delta \log(SFR)$ between adjacent time bins, with the Student's-t distribution used for the prior, following the approach in \citet{lej19a}. The total stellar mass form is another model parameter, which controls the normalization. I have tested various age bin configurations and found that the main conclusions of this paper are robust as long as sufficient time resolution is provided for the last few hundred Myrs. Additionally, I use a double power-law parametric form: $\psi(t) \propto [ (t/\tau)^\alpha + (t/\tau)^{-\beta} ]^{-1}$, to derive the SFH, and the main conclusions remain consistent. The stellar metallicity is allowed to vary between 0.01$Z_\odot$ and 2$Z_\odot$. 

To address the potential issue with spectrophotometric calibration, I incorporate a second-order Chebyshev polynomial perturbation and a white noise model into the spectroscopic data. This approach accounts for possible mismatch in the overall shape between the spectrum and the models, as well as any potential underestimation of the uncertainties. 

The emission-line model (Section~\ref{sec:ppxf}) is subtracted from the observed spectrum, leaving only the stellar continuum for model constraints. BAGPIPES employs a Bayesian approach to determine the most probable parameters. Table~\ref{tab:pipes_param} presents the priors and posteriors for the model parameters.

\begin{deluxetable*}{lllllccc}
\tablecaption{Priors and posteriors for the SFH fitting \label{tab:pipes_param}}
\tablehead{\colhead{Component} & \colhead{Parameter} & \colhead{Symbol / Unit} & \colhead{range} & \colhead{Prior} & \multicolumn{2}{c}{Hyperparameters} & \colhead{Posterior} }
\startdata
                & Redshift & z & (4.09, 4.12) & uniform & & & $4.10711^{+0.00026}_{-0.00023}$  \\
                & Stellar velocity dispersion & $\sigma$/(km/s) & (30, 500) & logarithmic & & & $330^{+17}_{-18}$ \\ 
			\hline
		SFH & Stellar mass formed & $\log(M_{\ast}/M_\odot)$ & (10, 12)  & uniform & & & $11.00^{+0.22}_{-0.14}$ \\
   & Current stellar mass\tablenotemark{a} & $\log(M_{\ast}/M_\odot)$ & & & & & $10.80^{+0.21}_{-0.14}$ \\
			& Stellar metallicity & $Z/Z_\odot$ & (0.01, 2) & logarithmic & & & $0.46^{+0.22}_{-0.14}$ \\
			\hline
		Dust & Dust attenuation at 5500\AA & $A_v$ / mag & (0, 5) & uniform & & & $0.44^{+0.45}_{-0.28}$ \\
    		& Deviation from Calzetti slope & $\delta$ & (-0.3, 0.3) & Gaussian & $\mu=0$ & $\sigma=0.1$ & $0.009^{+0.086}_{-0.077}$ \\
                & Birth cloud attenuation ratio & $\eta$ & (1, 4) & uniform & & & $2.68^{+0.79}_{-0.98}$ \\
			\hline
			Calibration & 0th order & & (0.75, 1.25) & Guassian & $\mu=1$ & $\sigma=0.25$ & $1.04^{+0.12}_{-0.15}$ \\
			& 1st order & & (-0.5, 0.5) & Guassian & $\mu=0$ & $\sigma=0.25$ & $0.12^{+0.09}_{-0.06}$ \\
			& 2nd order & & (-0.5, 0.5) & Guassian & $\mu=0$ & $\sigma=0.25$ & $-0.04^{+0.01}_{-0.01}$ \\
			\hline
			Noise model & White noise scaling & & (1, 10) & logarithmic & & & $1.92^{+0.04}_{-0.05}$  \\
\enddata
\tablenotetext{a}{The current stellar mass is not a fitting parameter but is calculated from the "stellar mass formed" parameter.}
\end{deluxetable*}

Figure~\ref{fig:gal}c illustrates the SFH of the quiescent galaxy. The galaxy formed half of its stellar mass $\sim670$~Myr ago, corresponding to $z\sim6.7$. The SFR peaked around 400~Myr ago, then decreased sharply from $\sim130 M_\odot$~yr$^{-1}$ to the current rate of $3 M_\odot$~yr$^{-1}$ -- a 40-fold decline, equivalent to an e-folding timescale of $\sim50$~Myrs in the past 200~Myrs. The present specific SFR (sSFR) is $5 \times 10^{-11}$~yr$^{-1}$, roughly 40 times lower than the star-forming main sequence at $z=4$ \citep{pop23}.

I also calculate the upper limit of current SFR from the H$\beta$ emission. After correcting for the dust attenuation as constrained by the stellar continuum (Table~\ref{tab:pipes_param}), including attenuation from birth clouds, the $2-\sigma$ upper limit of $SFR_{H\beta}$ is $0.6 M_\odot$~yr$^{-1}$, which is even lower than the SFR constrained by the stellar continuum and places SNH0pe-NS274 $\sim200$ times lower than the star-forming main sequence.

\section{Diagnostics of gas flow}
\label{sec:outflow}
The G140M grism spectrum provides coverage for various tracers for neutral gas in the restframe UV wavelength range. I use the ISM absorption spectrum, along with emission lines in the restframe optical (Section~\ref{sec:ppxf}), to quantify the physical properties of neutral and ionized gas in this quiescent galaxy. The same as in Section~\ref{sec:opt}, the quantities are reported without the slit loss applied. 

\subsection{Modeling the ISM absorption}
\label{sec:flow_model}

\begin{figure*}
\centering
\includegraphics[width=0.95\textwidth]{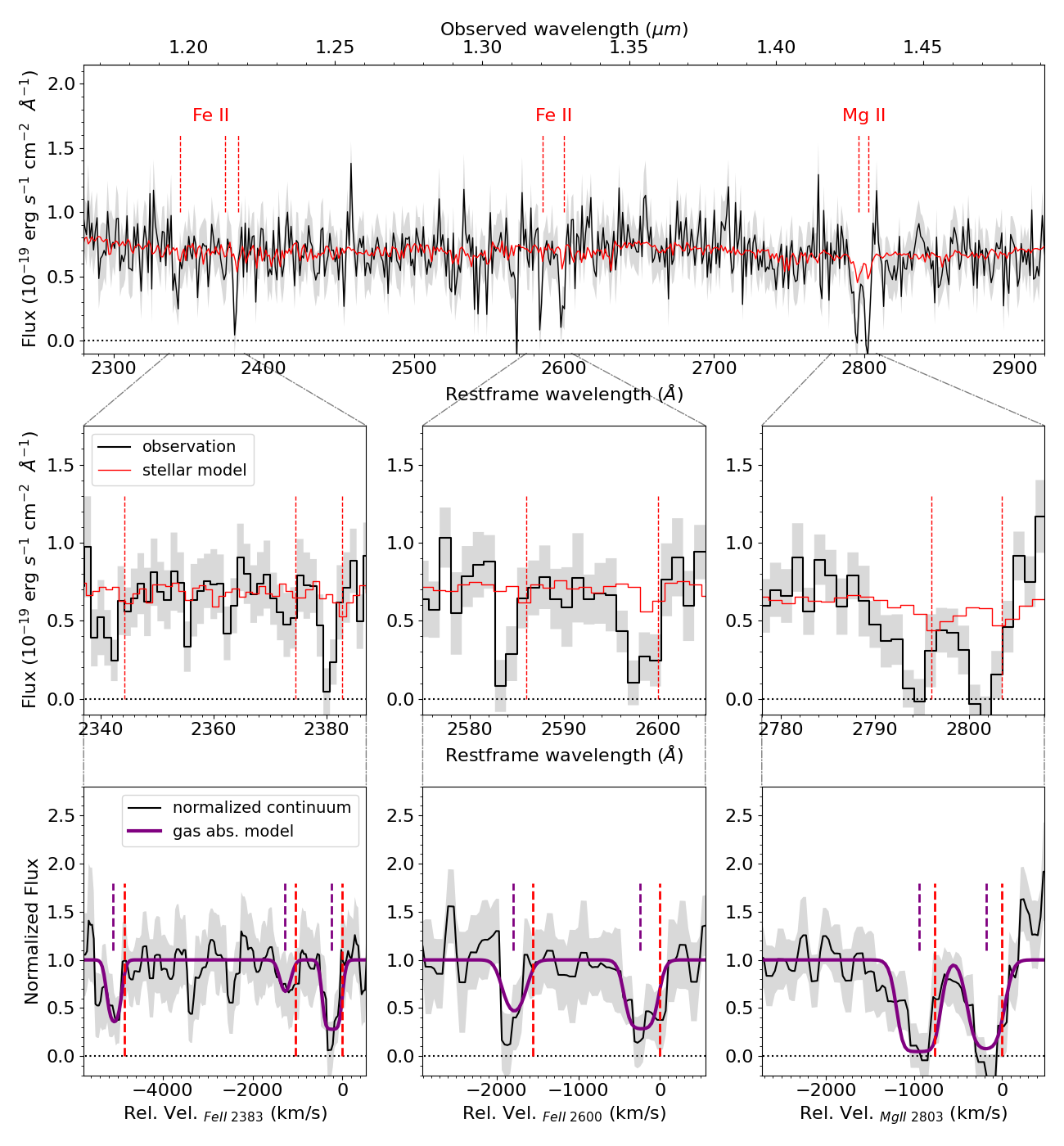}
\caption{\textit{Top:} The restframe UV spectrum (black) is shown with the model stellar continuum spectrum (red) derived from the best-fit parameters constrained by the restframe optical spectrum and normalized to the local continuum level. The spectrum features a series of \ion{Fe}{2} and \ion{Mg}{2} absorptions that are more pronounced than what would be expected from the stellar absorption alone. \textit{Middle:} A zoomed-in view of the absorption lines. The red dashed lines mark the wavelengths of absorption at rest based on the redshift determined from the optical spectrum. The centers of the absorption lines are blueshifted. The gray shade area is the uncertainty of the spectrum. \textit{Bottom:} The relative strength of the observed spectrum compared to the model stellar continuum. The thick purple lines represent the best-fit ISM absorption models. Red dashed lines indicate zero velocity for each absorption, while purple dashed lines are the best-fit line centers. Zero on the x-axis is set by the rest wavelength of the reddest line in each panel. Both \ion{Fe}{2} and \ion{Mg}{2} are blueshifted by $\sim200$~km~s$^{-1}$ relative to the stellar continuum. \label{fig:uv}}
\end{figure*}

The top panel of Figure~\ref{fig:uv} shows the restframe UV spectrum. Within this wavelength range, the most prominent absorption feature is the \ion{Mg}{2}$\lambda$2796,2803 doublet. Additionally, several \ion{Fe}{2} lines are clearly visible, including $\lambda$2344, 2383, 2586, and 2600. There is also a hint of the weaker \ion{Fe}{2}$\lambda$2374 line, although the absorption is consistent with the uncertainties. 

These UV absorptions may originate from both gas in the foreground and the atmosphere of A-type and later spectral type stars. The red line in Figure~\ref{fig:uv} represents the rescaled model of the stellar continuum, derived from the stellar population constraints obtained from the restframe optical spectrum (Section~\ref{sec:pipes}). The rescaling involves comparing the median flux within $\pm100$\AA\ of the observed and model spectra at each wavelength. 

The middle row of Figure~\ref{fig:uv} illustrates that the observed strengths of the observed \ion{Mg}{2} and \ion{Fe}{2} absorption lines exceed what can be explained by the stellar continuum alone. Therefore, the additional absorption is attributed to the foreground gas. {In addition, the absorption lines appear to be located at short wavelengths than expected positions.} 

The bottom row of Figure~\ref{fig:uv} shows the absorption attributed to the gas. The observed and model spectra are resampled onto a common 0.3~\AA\ grid in the restframe. The observed spectrum is then divided by the model continuum, resulting the black line shown. This division normalizes the spectrum to the stellar continuum. %Visual inspection suggests that the gas absorption features are blueshifted by $\sim200$~km~s$^{-1}$ relative to the stellar continuum. 

To model the gas absorption, I use the formulation from \citet{rup05}, for the  relative intensity as a function of wavelength, given by: $I(\lambda) = 1 - C_f + C_f e^{-\tau(\lambda)}$, where $C_f$ is the gas covering fraction. When $\tau(\lambda)=0$, $I(\lambda)=1$. As $\tau(\lambda)$ increases, $I(\lambda)$ approaches $1-C_f$ when $\tau(\lambda) \rightarrow \infty$. Given that the fluxes at the bottom of the \ion{Mg}{2}$\lambda 2796,2803$ and \ion{Fe}{2}$\lambda2383,2586,2600$ absorption lines are close to zero, it indicates that $C_f$ must be close to 1. 

The optical depth $\tau(\lambda)$ is modeled as a Gaussian: $\tau(\lambda) = \tau_0 e^{-(\lambda-\lambda_0)^2 / (\lambda_0 b_D / c)^2}$. $\tau_0$ is the optical depth at the line center.  $\lambda_0$ is the central wavelength of the line. The velocity of the line is then calculated from comparing $\lambda_0$ to the restframe wavelength of the line. $b_D$ is the Doppler parameter, which relates to the velocity width of the absorption. Finally, $c$ is the speed of light. 

While the covering fraction could in principle be a function of wavelength \citep{mar09}, it is assumed to be a constant here due to the lower spectral resolution. For a doublet, the combined intensity is given by $I(\lambda) = 1 - C_f + C_f e^{-\tau_1(\lambda)-\tau_2(\lambda)}$, where $\tau_1(\lambda)$ and $\tau_2(\lambda)$ are the optical depths for the two lines of the doublet. 

\begin{table}
\centering
\hspace*{-1.5cm}
\begin{threeparttable}
	\caption{Physical properties of \ion{Fe}{2} and \ion{Mg}{2} absorption lines and derived quantities}
\begin{tabular}{lcc}
	\hline
	\hline
           &  \ion{Fe}{2} & \ion{Mg}{2}  \\
        \hline
        (1) $v$ / km s$^{-1}$ & $-236^{+25}_{-20}$  & $-176^{+29}_{-28}$   \\
        (2) $C_f$ & $0.71^{+0.10}_{-0.11}$ & $0.98^{+0.02}_{-0.09}$  \\
        (3) $b_D$ / km s$^{-1}$ & $139^{+33}_{-26}$ & $148^{+27}_{-40}$ \\
        \hline
        \hline 
           &  \ion{Fe}{2}$\lambda2374$ & \ion{Mg}{2}$\lambda2803$  \\
        \hline
        (4) $\tau_0$ & $0.75^{+0.76}_{-0.33}$ & $4.7^{+7.8}_{-1.3}$ \\
        (5) $W/$\AA  & $0.81^{+0.25}_{-0.20}$ & $3.80^{+0.53}_{-0.58}$ \\
        (6) $\log[N(X)/\mbox{cm}^{-2}]$ & $14.89^{+0.15}_{-0.22}$ & $14.27^{+0.07}_{-0.08}$ \\
        (7) $\log[N(H)/\mbox{cm}^{-2}]$ & $21.21^{+0.15}_{-0.22}$ & $19.62^{+0.07}_{-0.08}$ \\
	\hline
\end{tabular}
\begin{tablenotes}
    \item (1) $v$: central velocity, calculated from the central wavelength of the absorption $\lambda_0$ (2) $C_f$: covering fraction. (3) $b_D$: Doppler parameter. (4) $\tau_0$: optical depth. (5) Equivalent width. (6) Column density of \ion{Fe}{2}$\lambda$2374 or \ion{Mg}{2}$\lambda$2803. (7) Hydrogen column density derived from \ion{Fe}{2}$\lambda$2374 or \ion{Mg}{2}$\lambda$2803.
\end{tablenotes}

\label{tab:abs}
\end{threeparttable}
\end{table}

The five \ion{Fe}{2} lines ($\lambda$2344, 2374, 2383, 2586, 2600) are modeled together as one single Doppler component. The ratios among $\tau_0$ of each \ion{Fe}{2} transition are fixed based on their oscillation strength \citep{mor03}. Similarly, the two \ion{Mg}{2} lines are modeled as one single Doppler component, and the ratio between the $\tau_0$ is fixed. A Gaussian noise is added to the model for 1,000 iterations to simulate the formal error spectrum and determine uncertainies of the model parameters (Table~\ref{tab:abs}). Both the \ion{Fe}{2} and \ion{Mg}{2} absorptions are found to be blueshifted by $\sim200$~km~s$^{-1}$ relative to the stellar continuum. The high covering fractions are consistent with the visual inspection (Figure~\ref{fig:uv}).

\subsection{Gas Column Density}

In the optically thin case, the column density of the gas can be estimated with the linear relation of the curve of growth \citep[e.g.][]{dra11}
\begin{equation}
N_{ion} [cm^{-2}] = 1.13 \times 10^{20} \frac{W_{ion}}{f \lambda^2} \frac{1}{C_f},
\end{equation}
where $W_{ion}$ is the equivalent width of the absorption line in angstrom, $f$ is the oscillator strength, and $\lambda$ is the wavelength of the line. 

The joint fit of the multiple \ion{Fe}{2} indicates that only the weakest \ion{Fe}{2}$\lambda$2374 line is optically thin ($\tau_0 = 0.75$, Table~\ref{tab:abs}). The second weakest line, \ion{Fe}{2}$\lambda$2586, has an oscillator strength 2.2 times that of \ion{Fe}{2}$\lambda$2374 thus $\tau_0 > 1$. Given this, I use the best-fit model of \ion{Fe}{2}$\lambda$2374 (thick red line in the bottom panels in Figure~\ref{fig:uv}) to estimate the gas column density. As for \ion{Mg}{2}, even the weaker \ion{Mg}{2}$\lambda$2803 line has $\tau_0 \gg 1$, therefore, only a lower limit of the column density can be obtained. 

I integrated the best-fit models over the wavelength ranges of the lines to calculate the equivalent widths. In many runs of fitting, the models suggest that \ion{Mg}{2}$\lambda$2803 is blended with \ion{Mg}{2}$\lambda$2796. In these cases, I take the wavelength with the highest flux between \ion{Mg}{2}$\lambda$2796 and $\lambda$2803 as the bluest wavelength and integrate redwards to obtain the equivalent width of \ion{Mg}{2}$\lambda$2803. The equivalent widths of \ion{Fe}{2}$\lambda$2374 and \ion{Mg}{2}$\lambda$2803 and the derived \ion{Fe}{2} and \ion{Mg}{2} column densities are listed in Table~\ref{tab:abs}.

Next, I use \ion{Fe}{2} to estimate the total hydrogen column density. This calculation requires an understanding of the ionization state of iron, the metallicity of the gas, and the degree of dust depletion for the elements. Current available data do not contain all the information, I thus adopt commonly used values and note that the derived hydrogen column densities depend on the assumptions.

Firstly, I assume that all Fe are singly ionized, $\chi = n(FeII)/n(Fe) = 1$. In \ion{Mg}{2} absorbing clouds at $z\sim1$, almost all Mg are singly ionized \citep{chu03}. Given that the ionization potentials of neutral Mg and Fe (7.6 eV and 7.9 eV) and \ion{Mg}{2} and \ion{Fe}{2} (15.0 eV and 16.2 eV) are similar, it is reasonable to assume that \ion{Fe}{2} is the dominant ion. 

Secondly, elements in the gas phase can be significantly depleted onto dust grains, and the depleted fraction varies widely depending on the composition and of the grains and other properties of ISM \citep{sav96}. The depletion factor $X$, denoted as $\delta(X)$, in defined as the logarithmic difference between the gas-phase abundance and the cosmic abundance: $ \delta(X) = \log(X/H)_g - \log(X/H)_0$. In the Galactic disk, the depletion factor for iron, $\delta(Fe)$, ranges between -1.0 and -2.3 \citep{jen09}. I adopt a median value of $\delta(Fe) = -1.7$ based on the sample measured by \citet{jen09}. Additionally, I use the solar abundance $\log (Fe/H) = -4.49$ \citep{sav96} and assume the gas-phase metallicity is the same as the stellar metallicity (as listed in Table~\ref{tab:pipes_param}). 

The hydrogen column density derived from \ion{Fe}{2} absorption (Table~\ref{tab:abs}) can thus be express as:
\begin{multline}
N(H) =  4.54\times10^{21}\ \mbox{cm}^{-2}  \left( \frac{N(FeII)}{10^{15}\  \mbox{cm}^{-2}} \right) \left( \frac{Z}{Z_\odot} \right)  \\ \left( \frac{10^{-4.49}}{10^{\log Fe/H}} \right) \left(\frac{10^{-1.7}}{10^{\delta(Fe)}} \right) \left( \frac{1}{\chi(FeII)} \right).
\end{multline}

Plugging in $N(FeII) = 10^{14.89}\ \mbox{cm}^{-2}$ (Table~\ref{tab:abs}) and $Z=0.46 Z_\odot$ (Tabel~\ref{tab:pipes_param}), I derive the hydrogen column density $N(H) = 1.62 \times 10^{21}\ \mbox{cm}^{-2}$. Table~\ref{tab:abs} also presents the lower limit of $N_H$ derived from \ion{Mg}{2}, assuming $\chi=1$, $\log(Mg/H)=-4.42$, and $\delta(Mg) = -0.8$, following the same references.

\subsection{Mass outflow rate}

The $\sim200$~km~s$^{-1}$ blueshifted \ion{Fe}{2} absorption and $\sim300$~km~s$^{-1}$ [\ion{O}{3}] emission indicate the presence of outflowing gas in both neutral and ionized phases. To estimate the amount of outflowing gas in each phase, I make a few reasonable and common assumptions.

\subsubsection{Neutral outflow from [\ion{Fe}{2}]}

Under a simple assumption of thin-shell geometry of the outflowing gas, the mass and rate of the outflowing gas can be estimated as follows:
\begin{subequations}
\begin{equation}
M_{out} = 1.4m_p \Omega N_H R^2_{out}
\end{equation}
\begin{equation}
\dot{M}_{out} = 1.4 m_p \Omega N_H R_{out} v_{out},
\end{equation}
\label{eq:mout}
\end{subequations}
where $m_p$ is the mass of the proton, $\Omega$ is the solid angle subtended by the outflow, $R_{out}$ is the radius of the shell, and $v_{out}$ is the velocity of the outflow  \citep{rup05}. Equation~\ref{eq:mout} can be expressed as:
\begin{subequations}
\begin{multline}
M_{out} = 1.3\times10^8 M_\odot \left( \frac{N_H}{10^{21} \mbox{cm}^2} \right)   
\left( \frac{\Omega}{4\pi} \right)
\left( \frac{R_{out}}{1 \mbox{kpc}} \right)^2
\end{multline}

\begin{multline}
\dot{M}_{out} = 36\ M_\odot \mbox{yr}^{-1} \left( \frac{N_H}{10^{21} \mbox{cm}^2} \right)  \\ 
\left( \frac{\Omega}{4\pi} \right)
\left( \frac{R_{out}}{1 \mbox{kpc}} \right) 
\left( \frac{v_{out}}{250\ \mbox{km s}^{-1} } \right)
\end{multline}
\end{subequations}

Neutral outflow are detected in $\sim45\%$ of massive ($M_\ast > 10^{10} M_\odot$) galaxies at $z\sim2$ \citep{dav24}. Assuming that these galaxies have outflows with a similar geometry and are oriented randomly, this result implies that $\Omega/4\pi > 0.45$. I thus adopt $\Omega = 0.45\times4\pi$. 

Given that the covering fraction $C_f$ of the outflow is close to 1 (Section~\ref{sec:flow_model}), the physical extent of the outflowing gas should at least match the extent of the stellar light. Therefore, I conservatively assume $R_{out} = 2 R_e = 780\ pc$. It is worth noting that kilo-parsec scale neutral outflows have been reported for a quiescent galaxy at $z\sim3$ \citep{deu23}.  

I then use the central velocity of the \ion{Fe}{2} absorption lines as the outflow velocity: $v_{out} = 236$~km~s$^{-1}$, which provides a conservative estimate by assuming a narrow cone of outflow directed towards the observer. 

Consequently, the mass of the neutral outflow and the outflow rate estimated from the \ion{Fe}{2}$\lambda2374$ line are $M_{out} = 6.0\times10^7 M_\odot$ and  $\dot{M}_{out} = 20\ M_\odot\ \mbox{yr}^{-1}$.
The outflow rate is approximately 7 times higher than the SFR inferred from stellar continuum and 30 times higher than the upper limit inferred from H$\beta$ emission (Section~\ref{sec:pipes}).

\subsubsection{Ionized outflow from [\ion{O}{3}]}

The mass of ionized gas in the outflow can be estimated from the [\ion{O}{3}] luminosity using the formula \citep{car15}:
\begin{equation}
    M_{out,[OIII]} = \frac{1.4 m_p L_{[OIII]}}{n_e (n_O/n_H)_\odot (Z/Z_\odot) j }, 
\end{equation}
where $m_p$ is the mass of the proton, $n_e$ is the electron density, and $j$ is the line emissivity. The term $(n_O/n_H)_\odot(Z/Z_\odot)$ is the element abundance, with $n_O/n_H = 10^{-3.13}$ \citep{sav96} and $Z/Z_\odot$ is taken from Table~\ref{tab:pipes_param}. Assuming a typical electron density of $n_e \simeq 500$~cm$^{-3}$ and a temperature of $T_e \simeq 10^4 K$ for the narrow-line region, the line emissivity is $j = 3.4 \times 10^{-21}$~ergs~s$^{-1}$~cm$^{-3}$ \citep{car15}. The mass outflow rate is then given by $\dot{M}_{out,[OIII]} = M_{out,[OIII]} \times v / R_{out}$. I adopt the central velocity of the [\ion{O}{3}] emission $v = 300$~km~s$^{-1}$ as the outflow velocity and an outflow radius $R_{out} = 2 R_e$. 

After correcting for the dust attenuation (Table~\ref{tab:pipes_param}), I derive the following values from the [\ion{O}{3}] emission: $L_{[OIII]} = 1.4 \times 10^{41}$~ergs~s$^{-1}$, $M_{ion,out} = 2.0\times10^5 M_\odot$ and $\dot{M}_{ion,out} = 0.08 M_\odot$~yr$^{-1}$. The ionized outflow rate is orders of magnitude lower than the neutral outflow rate.

\section{Discussion}
\label{sec:dis}

\subsection{Ejective feedback as a quenching mechanism}

Recently quenched galaxies --- quiescent galaxies that have experienced a recent, rapid decline in star formation ---, are observed across cosmic time. These galaxies show SFHs with e-folding timescales of $\sim100$~Myrs, far shorter than the gas depletion timescale of $\sim1$~Gyr typically found in star-forming galaxy \citep{tac18}. This disparity suggests that the formation of recently quenched galaxies cannot be explained simply by cutting off the cold gas supply. Several massive quiescent galaxies at $z>4$ have now been spectroscopically confirmed \citep{tan19,val20,car23,car24,fry24,kak24,deg24}. Given the young age of the universe at these redshifts, star formation in these galaxies must have been shut down on a very short timescale. Take SNH0pe-NS274 as an example: its e-folding time of 50~Myr is much shorter than the typical gas depletion time of $\sim500$~Myr for star-forming galaxies at $z\sim4$ \citep{des20}. Mechanisms capable of destroying the cold gas reservoir must have already been in action at $z>4$.

At lower redshifts, high-velocity outflows from recently quenched galaxies have been observed in several cases \citep{tre07,coi11,mal19,wu23,tay24}. These outflows, reaching velocities over 1000~km~s$^{-1}$, appear in either neutral phase ---  traced by \ion{Mg}{2} absorption --- or the ionized phase, marked by [\ion{O}{3}] emission, and can be driven by both AGN and starburst. However, the mass outflow rates in these galaxies remain unclear. \ion{Mg}{2} absorption, often optically thick, typically provides only a lower limit on the gas column density, while the ionized [\ion{O}{3}] gas traces only a minor fraction of the total gas mass \citep{fio17,bar20,ave22}. 

Compelling evidence for ejective feedback as a quenching mechanism comes from two massive recently quenched galaxies at $z=2.5$ and $z=3$ \citep{deu23,bel24}. The estimated outflow rates of neutral gas, traced by \ion{Na}{1}, exceed the SFRs by over 10 times. In both galaxies, the observed [\ion{N}{2}]/H$\alpha$ and [\ion{O}{3}]/H$\beta$ ratios indicate the presence of AGN, pointing to quenching driven by ejective feedback from supermassive black holes.

SNH0pe-NS274 presents a similar scenario at an even higher redshift. The neutral outflow rate is at least $\sim7$ times the SFR. This indicates that the gas reservoir is depleted mainly by the outflow rather than the star formation. Consequently, the SFR declines on a much shorter timescale than the typical gas depletion time observed in star-forming galaxies. Similar to lower-z galaxies, the ionized gas represents only a small fraction of the total mass ejected from SNH0pe-NS274. Despite that emission lines are easier to detect, measuring the mass budget from neutral phase outflows provides more direct evidence for ejective feedback as a quenching mechanism.

It is important to note that the measured outflow velocity of $\sim250$~km~s$^{-1}$ might not be sufficient to escape the gravitational potential of a massive galaxy halo. The gas could fall back and potentially fuel future star formation. Nevertheless, the case presented here underscores the capability of ejective feedback to trigger quenching, halt star formation at least temporarily, and help explain the presence of massive quiescent galaxies at $z>4$. 

\subsection{Can the outflow be driven by AGN?}
\label{sec:energy}

AGN is a theoretically promising mechanism for driving outflows, inducing ejective feedback, and ultimately shut down star formation. The high [\ion{O}{3}]/H$\beta$ and [\ion{Ne}{3}]/[\ion{O}{2}] ratios observed in SNH0pe-NS274 resemble those seen in AGN at $z>5$. If these emission lines are indeed powered by AGN, the strength of the [\ion{O}{3}] line can provide insights into the  energetics. Here, I explore whether the AGN-driven energetics are sufficient to plausibly drive the observed outflow.

Firstly, the bolometric luminosity of the AGN can be estimated as $L_{bol,AGN} = 600 L_{[OIII]}$, with a systemic uncertainty of a factor of 2 \citep{net09}. Numerical simulations suggest that only a small fraction, approximately $\sim5-15$ percent, of the bolometric luminosity is mechanically coupled to the ISM to produce realistic galaxies \citep{dim05,boo09,kur09}. For a conservative estimate, I adopt a lower value of 5\% and calculate the energy and momentum fluxes injected to the ISM from the AGN: $\dot{E}_{AGN} = 0.05 \times L_{bol,AGN} = 5.7\times 10^{42}$~erg~s$^{-1}$ and $\dot{p}_{AGN} = L_{AGN}/c = 3.8 \times 10^{35}$~dyne.

Next, the energy and momentum rates of the neutral outflow are calculated from the outflow rate and the outflow velocity: $E_{out} = \frac{1}{2} \dot{M}_{out} v_{out}^2 = 3.4 \times 10^{41}$~erg~s$^{-1}$ and $\dot{p}_{out} = \dot{M}_{out} v_{out} = 2.9 \times 10^{34}$~dyne. The ionized outflow is not considered here due to its negligible contribution to the total energy and momentum rates. 

The energy and momentum rates of the observed outflow are an order of magnitude smaller than the outputs from AGN. Although this estimation carries systematic uncertainties, it supports the scenario in which AGN could feasibly drive the outflow if it powers the observed [\ion{O}{3}] emission. Recently, studies have identified two quasars at $z>6$ situated in galaxies with low SFRs and rapidly declining SFHs \citep{ono24}. SNH0pe-NS274 may represent a subsequent stage of such systems, where the neutral outflow bridges the link between AGN and star formation quenching. This case highlights how AGN-driven feedback can play a pivotal role in suppressing star formation and shaping the evolution of massive galaxies in the early universe.

\section{Summary and Conclusion}
\label{sec:sum}
This paper reports the first detection of neutral gas outflow from a massive quiescent galaxy at $z>4$. Observed by JWST NIRSpec, the restframe optical spectrum of SNH0pe-NS274 shows no star-formation-related emission lines such as H$\beta$ and [\ion{O}{2}], while weak high-ionization lines of [\ion{Ne}{3}] and [\ion{O}{3}] present.

Modeling the stellar continuum absorption features yields a very low sSFR of $5 \times 10^{-11}$~yr$^{-1}$, roughly 40 times below the star-forming main sequence at a similar redshift. The non-detection of H$\beta$ emission puts an even more stringent constraint. Furthermore, the SFR of this galaxy has experienced a 40-fold decline over the last 400~Myr. Such a rapid truncation of star formation cannot be solely attributed to gas consumption. 

The restframe UV spectrum reveals blueshifted \ion{Mg}{2} and \ion{Fe}{2} absorption features at $\sim250$~km~s$^{-1}$ relative to the stellar continuum, signaling an outflow of neutral gas. The neutral outflow rate, estimated from the optically thin \ion{Fe}{2} line, is $\sim20 M_\odot \mbox{yr}^{-1}$ --- 7 times the SFR derived from stellar continuum and over 30 times the rate inferred from H$\beta$ emission. If the ionized emission line is AGN-driven, the energetics imply that the AGN can supply sufficient energy and momentum to drive the outflow. The high mass-loading factor in SNH0pe-NS274 offers compelling evidence that ejective feedback plays a critical role in quenching star formation in the early universe, shedding light on the presence of massive quiescent galaxies at $z>4$. 

\begin{acknowledgments}

I thank the referee who provided several constructive opinions, making the presentation better than the original form. I thank Prof. Lan Ting-Wen, Prof. Sirio Belli, Dr. Adam Carnall, Prof. David Rupke, and Prof. Misawa Toru for their suggestions and fruitful discussions. I acknowledge funding through the National Science and Technology Council grant 111-2112-M-002-048-MY2 and 113-2112-M-002-027-MY2. The data products presented herein were retrieved from the Dawn JWST Archive (DJA). DJA is an initiative of the Cosmic Dawn Center (DAWN), which is funded by the Danish National Research Foundation under grant DNRF140.
 
\end{acknowledgments}

\software{Astropy \citep{astropy13,astropy18,astropy22}, BAGPIPES \citep{car18,car19}, Galfit \citep{pen10}, pPXF \citep{cap04,cap17,cap23}, Matplotlib \citep{hun07}, Spectres \citep{car17}}

\bibliography{z4q_MgII}
\bibliographystyle{aasjournal}

\end{document}